# Identification of RF&MW microcalorimeter weak points by means of uncertainty analysis

Luca Oberto

*Abstract*—Microcalorimeters are used by National Metrology Institutes (NMI) for the realization of the Radiofrequency and Microwaves (RF&MW) primary power standard. Since they are not available on the market, NMIs have to design their own systems. It involves the design of complex mechanical structures and of low noise data acquisition systems. Resulting setups allow precision measurements of RF&MW power by applying difficult and time consuming measurement techniques. Weaknesses to be addressed in order to improve the performance, whether in the mechanical structure or in the data acquisition electronics, are not easy to identify. Anyway, they can be found by analyzing the microcalorimeter measurement uncertainty. In recent years, several improvements have been made on the INRIM (Istituto Nazionale di Ricerca Metrologica, Italy) coaxial microcalorimeter with the aim of making its operation more stable and to reduce the measurement uncertainty. Its weak points are here analyzed by means of a comprehensive uncertainty budget. Details are also given on how uncertainty contributions have been determined. The method here applied is general and can be conveniently used in many other fields or applications provided that the analytical model is known. It is useful while designing an experiment and/or *a posteriori*, to identify the parts of the apparatus and of the measurement set-up that need to be optimized.

## I. Introduction

THE Radiofrequency and Microwave (RF&MW) primary power standard is realized by measuring the effective efficiency of a suitable power sensor. This is made through microcalorimetric experiments [1]. The technique dates back to the early 50s and it is still unrivaled [2], [3].

Traditionally, bolometers are used for this application. Anyway, they are now substituted by thermoelectric sensors [4], [5], as bolometers cannot easily be found on the market. Thermoelectric sensors have also the advantage of not being downward frequency limited (i.e. their bandwidth extends down to dc) and of being less prone to environmental temperature variations [6]. Their effective efficiency is found with the same method used for bolometers, that involves thermal measurements with which the amount of RF&MW power lost into the sensor mount is determined.

Critical points of the microcalorimetric measurements are: the thermal stability of the measurement chamber, the stability of the RF&MW power and that of the reference power, the substitution error that occurs when an equivalent reference power is substituted to the RF&MW one, the measurement and connections repeatability.

Since microcalorimeters are complex systems and the related measurement techniques are time consuming, the identification of the weaknesses of those measurement systems is not an easy task. Anyway, it is mandatory for every National Metrology Institute willing to establish or improve a calibration service for RF&MW power. In particular, because microcalorimeters are not commercially available and have to be designed in house. A thorough evaluation of their measurement uncertainty is a key point to let weak points to emerge.

In recent years, several improvements have been made on the INRIM coaxial microcalorimeter. In this paper, we demonstrate how its most relevant critical point has been identified by means of uncertainty analysis. A comprehensive uncertainty budget is presented and, for each of the main uncertainty contributions, a description of the method used for its evaluation is given.

## II. Effective Efficiency Determination

For the determination of the effective efficiency of a suitable thermoelectric sensor mount, we apply a technique previously developed at INRIM [6-8]. It consists in supplying the RF&MW power to the sensor (at the 1 mW level) at the frequency of interest, record the asymptotic value $e_1$ of the temperature rise due to its losses, and then substitute the RF&MW power with a reference power that produces the same sensor output. The fall of the sensor temperature due to the absence of losses at the reference frequency (typically 1 kHz) is again recorded (asymptotic value $e_2$). Later, the sensor input is short-circuited and the whole measurement process is repeated. Short circuit measurements account for the coaxial feeding line losses.

The equation that governs the process is the following [9]:

$$\eta_e = \frac{e_2}{e_1} \cdot \left[1 - (1 + |\Gamma_S^2|) \cdot \frac{e_{1SC}}{2e_1}\right]^{-1} \cdot \frac{V_{U1}}{V_{U2}}, \quad (1)$$

L. Oberto is with the Nanoscience and Materials Division, Istituto Nazionale di Ricerca Metrologica (INRIM), Torino, 10135 Italy (e-mail: l.oberto@irim.it).

in which $e_{1SC}$ is the asymptotic value of the temperature reached when the short circuited sensor is fed with the RF&MW power and $\Gamma_S$ is the sensor reflection coefficient. $V_{U1}$ and $V_{U2}$ are the sensor output voltages when it is fed with the RF&MW and the reference power, respectively.

There exists another expression for the effective efficiency that takes into account also the reflection coefficient of the short circuit but, in our case, this contribution turned out to be negligible [9].

The technique here described rely on the assumption that the losses at the reference frequency are negligible, an assumption not always easy to verify. However, a new kind of thermoelectric sensors have been recently introduced on the market. They allow to relax this hypothesis by means of a second auxiliary absorber, coupled to the RF&MW one, that can be fed with the reference power through an alternative input [10]. We demonstrated that their effective efficiency can be evaluated with the same technique here described, that (1) still holds and that the hypothesis of absence of losses at the reference frequency is well applicable to the INRIM microcalorimeter when sensors with no auxiliary heater are used [11].

## III. UNCERTAINTY COMPONENTS

Recently, an evaluation of the uncertainty for these measurements has been made [12], but some important components were not analyzed. Indeed, a thorough evaluation is critical for the identification of the weaknesses of the microcalorimetric system. In this section and in the following section IV, a comprehensive analysis is carried out. All the relevant uncertainty sources are discussed and the procedure used to find their contribution to the overall uncertainty is explained.

### A. Uncertainty from the fitting process

The values of the parameters $e_1$, $e_2$ and $e_{1SC}$ has been calculated by fitting with and exponential curve the temperature trend, that follows the heating and cooling of the sensor mount and the coaxial feeding line when they are fed with RF&MW or LF reference power.

The temperature has been measured by means of a thermopile system that senses the change of the sensor temperature at the base of its connector. A nano-voltmeter was used to measure the thermopile voltage. Its instrumental uncertainty has been taken into account in the fitting process. It gives, as a result, the values of the asymptotic parameters we are interested in, along with their uncertainty $u_{fit}$.

### B. $e_1$, $e_2$ and $e_{1SC}$ measurement repeatability

Measurement repeatability $u_{rep}$ has been evaluated by performing several measurement cycles and evaluating the asymptotic values of the temperature trend with a fitting procedure for every cycle.

In Fig. 1 an example of repeated measurements is shown. The different curves are well superimposing and almost indistinguishable in the graph. The standard uncertainty arising

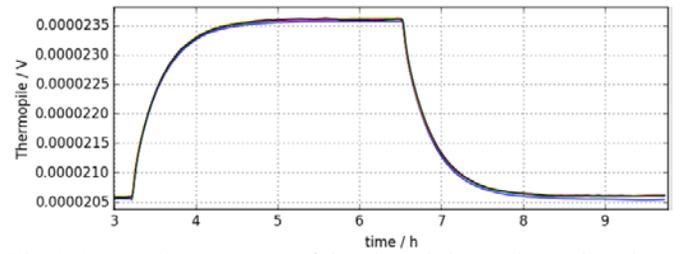

Fig. 1. Repeated measurements of the microcalorimeter thermopile voltage when the device under test (DUT) is heated by RF&MW losses and when it cools down after the power substitution with the reference 1 kHz power. (Color online)

from the repeated measurements for the parameters $e_1$, $e_2$ and $e_{1SC}$ turned out to be about 7 nV.

Of course, the measured temperature depends upon the stability of the measurement chamber temperature. The electronics of the microcalorimeter thermostat has been improved so that its peak-to-peak oscillation falls within the milliKelvin range, which gives a negligible contribution. See section III.F.

### C. Connection repeatability

This contribution takes into account both the repeatability of the connection of the sensor to the coaxial feeding line and the connection of the thermopile system that sense the temperature change to the sensor itself. The evaluation has been carried out by opening the microcalorimeter, removing the sensor and placing it in place again 10 times. After a proper thermal stabilization, the measurement procedure has been repeated and the asymptotic value of the thermopile reading has been acquired as in section III.A. The standard deviation of this values has been taken as an evaluation of the connection repeatability $u_{conn}$.

### D. Power sensor reflection coefficient

In (1), the reflection coefficient $\Gamma_S$ of the power sensor under characterization is required. It has been measured by means of a Vector Network Analyzer (VNA). The measurement uncertainty $u_{VNA}$ has been evaluated by means of the software VNA Tools II developed by the Swiss National Metrology Institute METAS [13]. The software requires the input of typical manufacturer data related to the component used by the VNA system, therefore this uncertainty contribution is considered as type B.

The realization of the removable short circuit used in the measurement process can be tricky, and obtaining a high reflection coefficient for this component is not easy, especially at frequency higher than 18 GHz. As a consequence, the reflection coefficient of the short circuited sensor $\Gamma_{SC}$ may be needed for the evaluation of the sensor effective efficiency [9]. It can be measured easily with a VNA but, in our case, it turned out to be worthless. Indeed, it appears in the equation in the following form [7]:

$$\eta_e = \frac{e_2}{e_1 - e_{1SC} + \frac{e_{2SC}}{|\Gamma_{SC}|^2}}, \qquad (2)$$

where $e_{2SC}$ is related to the losses of the feeding line and short circuited sensor mount at the reference frequency. Equation (2) is simplified with respect to (1): it is valid only if half of the power is supplied while in short circuit conditions. Moreover, the corrective terms containing $\Gamma_S$, $V_{U1}$ and $V_{U2}$ have been dropped for seek of simplicity.

By careful design, and after extensive experimental confirmation, the term $e_{2SC}$ is almost null, leading to a negligible contribution of the short circuit reflection coefficient.

Anyway, by using sensors equipped with auxiliary heater for the reference power [11], this term in the equation is cancelled by the simple fact that the reference power can be supplied to the sensor through an alternative input instead of through the coaxial feeding line whose losses we need to assess.

*E. Substitution error*

Substitution error is what arises when the reference power value is not equal to that of the RF&MW power to which it is substituted. It is recognizable by the fact that the two power sensor outputs are not equal within the measurement uncertainty. This systematic contribution is taken into account in (1) by means of the ratio of the two voltages $V_{U1}$ and $V_{U2}$ that are the sensor output voltages when it is supplied with the RF&MW power and with the reference power, respectively.

In the past, we evaluated for our system a substitution error of the order of 10 µW peak-to-peak. It was clearly noticeable in the measurements (see, for example, Fig. 2); therefore, its contribution could not be neglected [5]. This was due to the insufficient resolution of the generators and to their stability over time (a typical sensor characterization with the microcalorimeter usually requires a day per frequency, therefore a typical measurement session can last for two months).

To reduce this contribution, two ΔΣ modulators driven by two PID controllers have been introduced, that modulates the reference and the RF&MW power separately, so that the difference among the two average values is kept as constant and as low as possible. The modulators and PIDs are software implemented into the measurement software. As a result, the peak-to-peak oscillation of the power (both RF&MW and reference) has been reduced to 100 nW and no substitution error

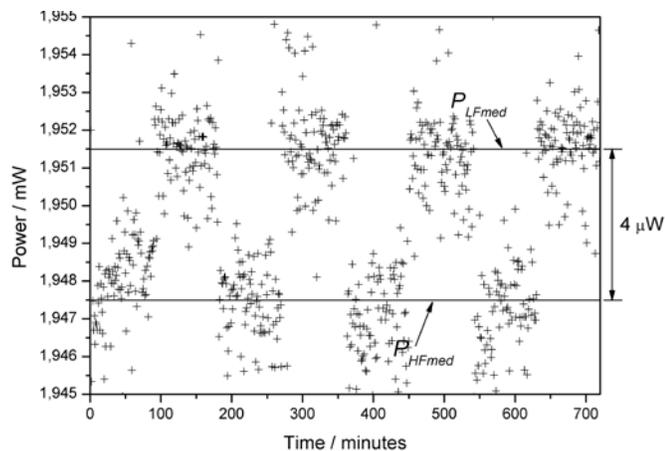

Fig. 2. Typical power stability over time. In this graph, the power is switched between RF and reference power every 90 minutes starting with RF. $P_{LFmed}$ and $P_{HFmed}$ represent the mean value of the reference and RF&MW power, respectively.

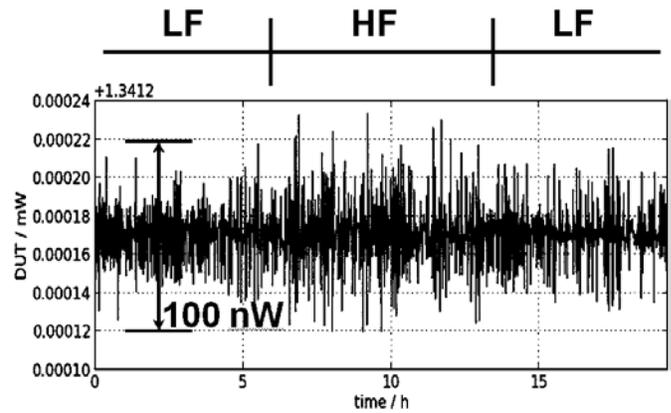

Fig. 3. A typical power substitution after the stabilization. As can be seen, no significant substitution error occurs. DUT stands for Device Under Test. LF stands for reference power while HF stands for RF&MW power. Here the power is switched between LF and HF every 6 hours.

is now visible on the data. Therefore, this uncertainty contribution can be neglected.

In Fig. 3 a typical power substitution obtained after the stabilization is shown. As can be seen by comparing Fig. 2 with Fig. 3, both power stability and substitution error are much improved by the modulation. Both Fig. 2 and Fig. 3 refer to the same microwave frequency of 18 GHz.

As previously said, in (1) the substitution error is taken into account by the ratio $V_{U1}/V_{U2}$. By neglecting this contribution, equation (1) is simplified as follows:

$$\eta_e = \frac{e_2}{e_1} \cdot \left[1 - (1 + |\Gamma_S^2|) \cdot \frac{e_{1SC}}{2e_1}\right]^{-1}. \quad (3)$$

*F. Thermal stabilization*

When measuring small temperature variations like the ones that the sensor under calibration undergoes in the microcalorimeter, the temperature stability of the measurement chamber is of utmost importance [14].

The last version of the INRIM microcalorimeter is composed by three metallic shields, the external one being passive, the intermediate being active by means of Peltier elements, while a wire heater is winded up around the internal shield. The two active shields are thermally stabilized by means of separate PID controllers. The space between the shields is filled with insulating foam. All the system is encased by insulating materials, overstepped only by copper fingers that connect the Peltier elements to their heat exchangers. See, for example, Fig. 4 taken from [14]. Furthermore, the microcalorimeter is placed in a shielded room whose temperature is stabilized in the ± 0.5 K range while the humidity is kept constant within ± 5 %.

In this way, a thermal stability of the measurement chamber of the order of 1 mK peak-to-peak has been reached. This can be maintained for a period as long as two months. Fig. 5 shows an example of the temperature behavior during about 20 hours. Once reached this thermal stability, its contribution to the sensor temperature measurement made with the thermopile assembly of the microcalorimeter is negligible.

In Fig. 6 a block diagram of the microcalorimeter electronics is shown. The PID controller for the thermal stabilization are highlighted along with the ΔΣ modulators and related PID controllers for the RF&MW and reference power stabilization.

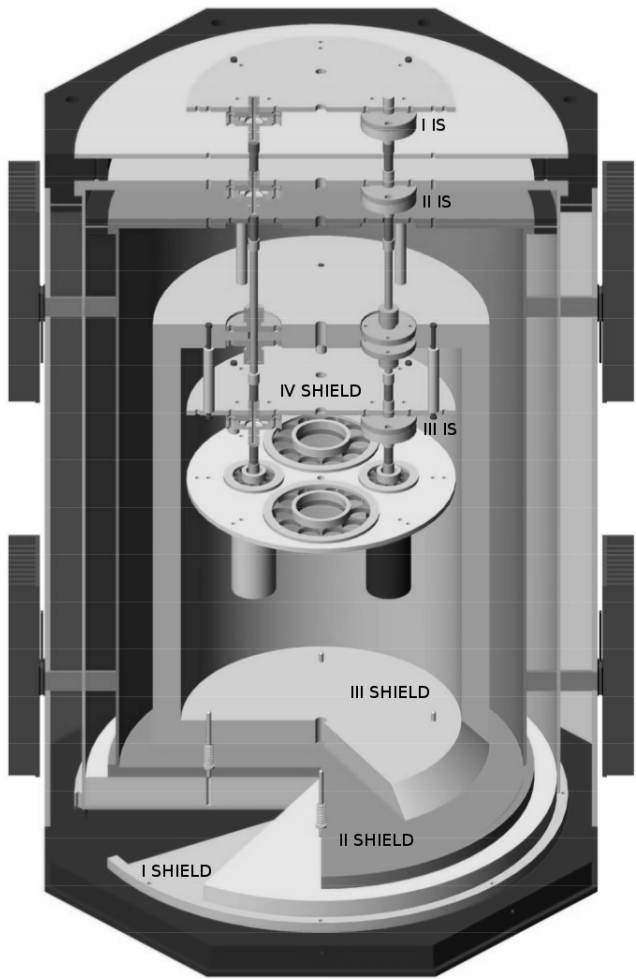

Fig. 4. A CAD picture of the INRIM microcalorimeter. Two sensors connected to the twin coaxial feeding lines are also represented. ISs are coaxial insulating sections. This picture is taken from [14].

$\Delta\Sigma$ modulators [15] have been necessary to overcome the insufficient power resolution of the RF&MW and LF synthesizers. They alternate every 0.5 s, much faster than the system time constant (about 2400 s), two adjacent values of power in such a manner that the average value coincides with the desired one.

### G. Power sensor linearity

The linearity of the power sensor is not an issue in the case of the power standard realization by means of the microcalorimetric technique, because the measurements are made at a constant power level, as Fig. 3 shows.

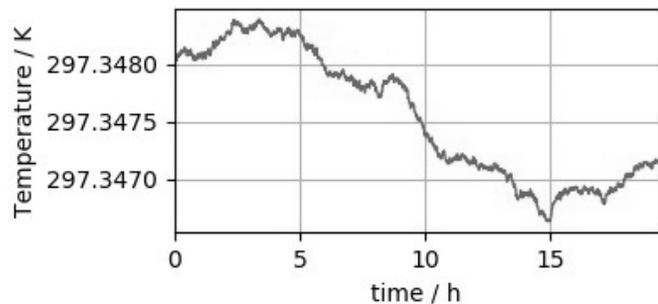

Fig. 5. Example of temperature measurement inside the measurement chamber of the INRIM microcalorimeter during 16 hours.

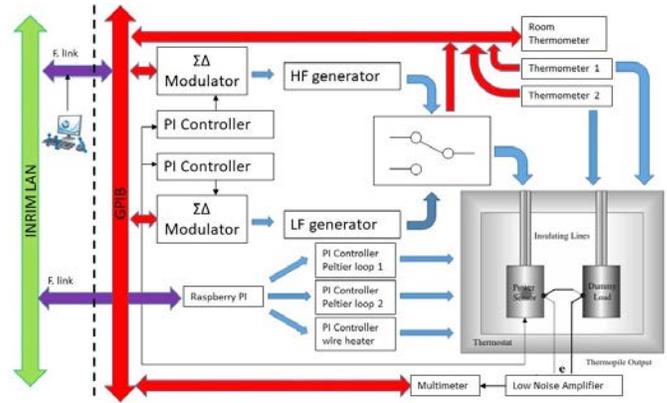

Fig. 6. Block diagram of the microcalorimeter electronics. PID controllers for both thermal and power stabilization are shown as well as the $\Delta\Sigma$ modulators. What is on the right side of the dashed vertical line is inside the shielded room. (Color online)

### H. Linearity of the microcalorimeter thermopile thermometer

As already stated, a thermopile thermometer is used to sense the temperature variation of the power sensor under calibrations and of the coaxial feeding line assembly. This temperature variation is due to the losses at RF&MW. Since the losses increase with frequency, the linearity of the thermopile, if not checked, may cause some issues and an extra uncertainty component. To verify if any linearity error is present, the thermopile voltage asymptotic value has been recorded (at a fixed frequency) for several RF&MW power levels, covering a broader range with respect to the actual power variations. Measurements have been taken at 0.25, 0.5, 1, 1.5 and 2 mW whilst, in practice, the sensor characterization is performed at a fixed, not critical, point around 1 mW with very small variations as shown in Fig. 3.

No meaningful deviation from linearity has been found within the measurement uncertainty. Results are shown in Fig. 7.

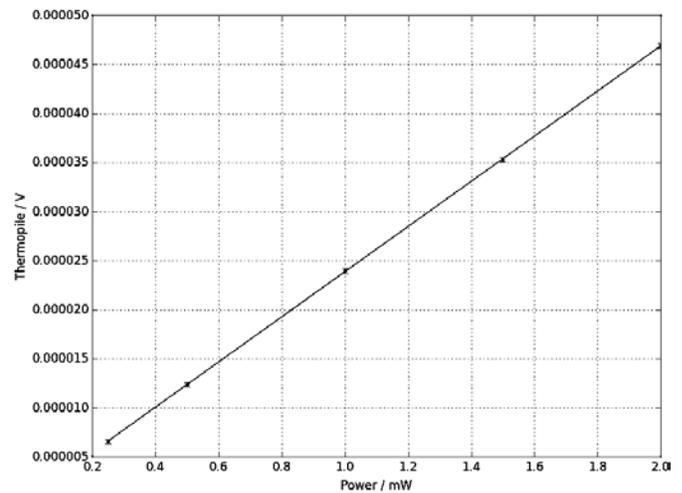

Fig. 7. Evaluation of the linearity of the microcalorimeter thermopile thermometer. No meaningful deviation from linearity has been found within the measurement uncertainty (uncertainty bars are present in the graph, but not easily visible being very small).

## IV. UNCERTAINTY CONTRIBUTIONS EVALUATION

The contribution of each uncertainty component described in Section III has been evaluated by means of the Law of Propagation of Uncertainty (LPU) [16] applied to (3). This led to an uncertainty budget that is presented in this section.

In the following, equations for every contribution are given. Their value turned out to be almost independent of the measurement frequency, except the term related to the power sensor reflection coefficient which, anyway, turned out to be the less influent.

### A. Uncertainty contribution arising from the measurement of the power lost in the system at RF&MW

The term in (3) that accounts for the RF&MW losses in the sensor mount is $e_1$. It has been evaluated by fitting the thermopile voltage output with an exponential curve. Three uncertainty components characterize this term, that are: fitting, measurement repeatability and connections repeatability. The former being evaluated as described in section III.A, the second as in section III.B and the latter as in section III.C.

According to the LPU, its contribution is given by:

$$u_{e1} = \sqrt{\left\{\frac{-4e_2}{[2e_1-(1+|\Gamma_S|^2)e_{1SC}]^2}\right\}^2 \cdot \left(u_{fit\_e1}^2 + u_{rep\_e1}^2 + u_{conn\_e1}^2\right)} \quad (4)$$

### B. Uncertainty contribution arising from the measurement of the power loss in the system at the reference frequency

This contribution is represented in (3) by the term $e_2$. It is evaluated, as the previous term $e_1$, by fitting the thermopile voltage output with a decreasing exponential curve. In fact, at the reference frequency, the losses in the feeding line and in the sensor mount are negligible, hence the temperature decreases after the substitution of the RF&MW power with the reference one. This term is affected by the same three uncertainty components that characterize $e_1$.

Therefore, its expression is given by:

$$u_{e2} = \sqrt{\left\{\frac{2}{2e_1-(1+|\Gamma_S|^2)e_{1SC}}\right\}^2 \cdot \left(u_{fit\_e2}^2 + u_{rep\_e2}^2 + u_{conn\_e2}^2\right)} \quad (5)$$

### C. Uncertainty contribution arising from the measurement of the power loss in the coaxial feeding line at RF&MW

The temperature rise due to the losses in the coaxial feeding line needs to be separated by the ones due to the sensor mount. This is the reason of the measurements with the short circuited power sensor. Losses in the feeding line are taken into account in (3) by the term $e_{1SC}$. It is evaluated in the same way as the previous two terms $e_1$ and $e_2$ and it is characterized by the same three uncertainty sources.

Its expression is given by:

$$u_{e1SC} = \sqrt{\left\{\frac{2e_1(1+|\Gamma_S|^2)}{[2e_1-(1+|\Gamma_S|^2)e_{1SC}]^2}\right\}^2 \cdot \left(u_{fit\_e1SC}^2 + u_{rep\_e1SC}^2 + u_{conn\_e1SC}^2\right)} \quad (6)$$

### D. Uncertainty contribution arising from the measurement of the power sensor reflection coefficient

This contribution is represented in (3) by the term $\Gamma_S$. It is evaluated, as described in section III.D, by means of VNA measurements. It is considered of type B having a rectangular distribution, according to [16].

Its expression is given by:

$$u_{\Gamma S} = \sqrt{\left\{\frac{4e_2 e_{1SC}|\Gamma_S|}{[2e_1-(1+|\Gamma_S|^2)e_{1SC}]^2}\right\}^2 \cdot \left(\frac{u_{VNA}}{\sqrt{3}}\right)^2} \quad (7)$$

## V. DISCUSSION

Results of the evaluation performed according to the previous sections IV.A to IV.D are summarized in Table 1. The overall uncertainty has been evaluated as follows:

$$u(\eta_e) = \sqrt{u_{e1}^2 + u_{e2}^2 + u_{e1SC}^2 + u_{\Gamma S}^2}. \quad (8)$$

The most relevant uncertainty sources are $e_1$ and $e_2$, that contribute almost equally to the overall uncertainty. Furthermore, it is worth noting that the measurement uncertainty of $e_1$, $e_2$ and $e_{1SC}$ is the same.

The reason for that can be understood if we rearrange equations (4) to (7) to highlight the effect of the single uncertainty sources. In this way we obtain the following three expressions:

$$u_j = \sqrt{\left\{\frac{-4e_2}{[2e_1-(1+|\Gamma_S|^2)e_{1SC}]^2}\right\}^2 \cdot u_{j\_e1}^2 + \left\{\frac{2}{2e_1-(1+|\Gamma_S|^2)e_{1SC}}\right\}^2 \cdot u_{j\_e2}^2 + \left\{\frac{2e_1(1+|\Gamma_S|^2)}{[2e_1-(1+|\Gamma_S|^2)e_{1SC}]^2}\right\}^2 \cdot u_{j\_e1SC}^2}, \quad (9)$$

in which $j = fit, rep, conn$, thus indicating the three uncertainty sources $u_{fit}$, $u_{rep}$ and $u_{conn}$ as defined in sections III.A, III.B and III.C. The fourth term we still need is just $u_{\Gamma S}$ as defined in (7).

TABLE I
UNCERTAINTY BUDGET FOR THE EVALUATION OF THE EFFECTIVE EFFICIENCY OF A THERMOELECTRIC POWER SENSOR FITTED WITH PCN CONNECTOR AT 18 GHZ. $\Gamma_s$ AND $\eta_e$ ARE DIMENSIONLESS, OTHER QUANTITIES AND THEIR UNCERTAINTIES ARE IN VOLT.

| Parameter | Value | Meas. Unc. | Sensitivity coeff. | Unc. Contrib. |
|---|---|---|---|---|
| $e_1$ | 2.5873×10⁻⁵ | 1.20216×10⁻⁷ | 3.98×10⁴ | 4.78×10⁻³ |
| $e_2$ | 2.2943×10⁻⁵ | 1.20213×10⁻⁷ | 4.16×10⁴ | 5.00×10⁻³ |
| $e_{1SC}$ | 0.3698×10⁻⁵ | 1.20223×10⁻⁷ | 1.99×10⁴ | 2.39×10⁻³ |
| $\Gamma_S$ | 0.0174 | 0.0080 | 2.56×10⁻³ | 2.07×10⁻⁵ |
| $\eta_e$ according to (3) | | | | 0.9550 |
| Overall uncertainty $u(\eta_e)$ | | | | 0.0073 |
| Expanded Unc. ($k = 2$) $U(\eta_e)$ | | | | 0.0146 |

The overall uncertainty is evaluated as follows:

$$u(\eta_e) = \sqrt{u_{fit}^2 + u_{rep}^2 + u_{rep}^2 + u_{\Gamma S}^2}. \qquad (10)$$

We can, now, express the uncertainty budget in a different form as shown in Table 2. In this way we note that the overall uncertainty is largely dominated by connections repeatability that affect all $e_1$, $e_2$ and $e_{1SC}$ terms. Their uncertainty in (4), (5) and (6) is, therefore, dominated by the connections repeatability term $u_{conn}$, the fitting and measurement repeatability terms $u_{fit}$ and $u_{rep}$ being two order of magnitude lower, at least.

As described in section III.C, the connection repeatability term should be only due to the effect of the assembly and disassembly of the system but, during data analysis, another contribution appeared. In fact, looking at the temperature stability of the shielded room in which the microcalorimeter in placed, we noted that the ambient conditioning system that should maintain the temperature of the room in the range of ± 0.5 K around the set-point, underwent a technical fault which has still to be investigated. Due to this, the temperature stability was not as high as required.

By looking, then, at the correlation between room temperature and thermopile thermometer voltage for every measurement of the series, we noticed a strong correlation (Pearson's coefficient $\rho = 0.86$). On the counterpart, no significant correlation has been observed among room temperature and measurement chamber temperature and among measurement chamber temperature and thermopile voltage.

This means that, even if the temperature of the microcalorimeter measurement chamber is well stabilized and its residual oscillations does not have any impact on the thermopile signal, the power sensor is exposed to a temperature variation anyway.

Hence, when the room temperature exhibits too wide oscillations, there is still a residual heat flow entering the system through the coaxial feeding lines that the coaxial insulating sections are not able to suppress.

It is clear, from this analysis, that the most important action to be taken in order to improve the performance of the INRIM microcalorimeter is to re-design this part of the system in order to ensure a better rejection of the external temperature fluctuation.

It could be objected that the effect of the room temperature variation should be visible on the measurement repeatability described in section III.B too. In practice this is not the case because, when the measurements required for its evaluation were performed, the air conditioning system of the shielded room did not have any problem. Furthermore, in the evaluation of the contribution of III.B, no reconnection is involved.

Therefore, the contribution of the connections we obtained in this evaluation was due to both temperature and connections themselves. As a consequence, the system that performs the contact of the thermopile to the sensor mount should also be re-designed in order to lower its contribution.

An alternative interpretation is the following: the effect of the room temperature oscillations that still reach the measurement chamber through the coaxial feeding lines bypassing the thermal shields and insulating sections, should be suppressed by the differential configuration of the thermopile thermometer. In fact, it senses the temperature of the two feeding lines and removes the common-mode. The outer conductor of both the feeding lines is thermalized onto the shields. If, for any reasons, after the repeatability measurements of section III.B, something happened to the inner conductor of the feeding line that acts as a thermal reference so that it was interrupted, this could cancel the common-mode suppression acted by the thermopile differential configuration.

If this is the case, no re-design of the insulating section will be necessary. This point needs further investigation.

## VI. CONCLUSION

The performance of the INRIM microcalorimeter for the realization of the RF&MW primary power standard has been improved in the last years, especially from the electronics point of view. In this work, a comprehensive evaluation of its measurement uncertainty has been carried out in order to recognize what are the most relevant contributions that still needs to be addressed.

All uncertainty contributions have been here described along with the methods used for their determination.

Finally, an uncertainty budget has been formulated and, by analyzing its figures, the most important contributions have been highlighted.

By expressing the uncertainty budget in two different ways, it was found that the measurement uncertainty is largely dominated by the connection repeatability contribution. It was supposed to be only due to connections and reconnections of the thermopile to the power sensor under calibration. In practice, while performing the very time consuming measurements needed for its evaluation, we noticed that, besides the connection effect itself, there was a significant correlation among the thermopile thermometer signal and the room temperature while the shielded room air conditioning system was undergoing a failure and was not able to maintain stable the ambient temperature to the level required for these measurements.

This could be due to an insufficient thermal isolation of the coaxial insulating lines when the room temperature oscillates too much, or to a failure of the differential configuration that suppresses the effect of the heat flux that still enters the microcalorimeter measurement chamber through the coaxial feeding lines.

TABLE II
BREAKDOWN OF THE UNCERTAINTY SOURCES FOR THE DETERMINATION OF $\eta_e$

| Uncertainty source | Uncertainty Contribution |
| --- | --- |
| Fitting ($u_{fit}$) | $9.20 \times 10^{-5}$ |
| Repeatability ($u_{rep}$) | $4.26 \times 10^{-4}$ |
| Connections ($u_{conn}$) | $7.31 \times 10^{-3}$ |
| VNA ($u_{\Gamma S}$) | $2.07 \times 10^{-5}$ |
| Overall uncertainty $u(\eta_e)$ | 0.0073 |
| Expanded Uncertainty ($k = 2$) $U(\eta_e)$ | 0.0146 |

In the first case, both the thermopile connection system and coaxial insulating sections needs to be re-designed whilst, in the second, a damage to the coaxial feeding line that acts as a thermal reference likely happened.

Further investigations are needed but, in any case, it is now clear that the weakest part of the INRIM microcalorimeter is the mechanical one instead of the electronic one.

It is worth noting that the analysis the uncertainty components, here used to identify the weak point of the INRIM microcalorimeter, can be conveniently applied in every experiment in which the analytical model is known, regardless the scientific field or the specific application.

It can also be used *a priori* to foresee the achievable measurement uncertainty, in order to properly design an experiment, and, then, *a posteriori* to find out the parts of the experimental apparatus that need to be improved in order to reduce the final overall uncertainty, as done in this work.

Moreover, since the thermopile assembly and the coaxial feeding lines, composed by thermally insulating and non-insulating sections, are key components of every coaxial microcalorimeter, the procedure highlighted in this work applies not only to the INRIM microcalorimeter and may be useful to other NMIs too, to check their systems. In fact, the identification of such kind of weaknesses is really not trivial.

ACKNOWLEDGMENT

The author thanks his colleague C. E. Calosso for useful discussions and for the help in the software implementation of the ΔΣ modulators.

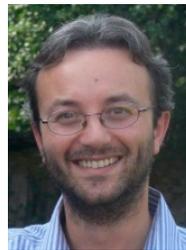

**Luca Oberto** (M'76–SM'81–F'87) was born in Pinerolo (Torino), Italy, on June 9, 1975. He received the M.S. degree in Physics from the University of Torino in 2003 and the Ph.D. in Metrology from the Politecnico di Torino in 2008.

From 2002 to 2003 he was with the Istituto Nazionale di Fisica Nucleare (INFN), Torino Section, working at the COMPASS experiment at CERN, Geneva, Switzerland. From 2003 he is with the Istituto Nazionale di Ricerca Metrologica (INRIM), Torino, Italy. His research interests are in the field of high frequency and THz metrology and in the realization and characterization of superconductive microwave devices.

Dr. Oberto is member of the Associazione Italiana Gruppo di Misure Elettriche ed Elettroniche (GMEE). He was recipient of the 2008 Conference on Precision Electromagnetic Measurements Early Career Award and of the GMEE 2010 "Carlo Offelli" Ph.D. prize.